\documentclass[aps,prd,twocolumn,showpacs,preprintnumbers,
amsmath,amssymb]{revtex4}
\usepackage{graphicx}
\begin{document}

\preprint{IHEP 2005--5}

\title{Synthetic Running Coupling of QCD}

\author{Aleksey I. Alekseev}
\email{alekseev@ihep.ru}
\affiliation{Institute for High Energy Physics, 142281 Protvino,
Moscow Region, Russia}


\begin{abstract}
Based on a study of the analytic running coupling obtained from
the standard perturbation theory results up to four-loop order,
the QCD ``synthetic'' running coupling  $\alpha_\mathrm{syn}$ is
built.  In so doing the perturbative time-like discontinuity is
preserved and nonperturbative contributions not only remove the
nonphysical singularities of the perturbation theory in the
infrared region but also decrease rapidly in the ultraviolet
region. In the framework of the approach, on the one hand, the
running coupling  is enhanced at zero and, on the other hand, the
dynamical gluon mass $m_g$ arises. Fixing the parameter which
characterize the infrared enhancement corresponding to the string
tension $\sigma$ and normalization,
say, at $M_\tau$ completely define the synthetic running coupling.
In this case the dynamical gluon mass  appears to be fixed and the
higher loop stabilization property of  $m_g$ is observed. For
$\sigma=$ (0.42 GeV)$^2$ and $\alpha_\mathrm{syn}(M^2_\tau)=$ 0.33
$\pm 0.01$ it is obtained that $m_g=$ 530 $\pm 80$ MeV.
\end{abstract}


\maketitle

\section{\label{sec1}Introduction}
The paper is devoted to description  of the running coupling (the
invariant charge) of QCD in which on the basis of the perturbative
study up to four-loop order the attempt has been undertaken   to
include the nonperturbative contributions in such a way that
nonphysical singularities of the perturbation theory in the
infrared region be cancelled and the essentials of the QCD
dynamics for all energy scales  accumulated in its framework.

Doing that it seems notably attractive,  on the theoretical
grounds of the analytic running coupling of QCD, to build the
running coupling including additional nonperturbative terms. We
will consider a variation of such couplings, the ``synthetic''
running coupling $\alpha_\mathrm{syn}(Q^2)$, which on the
practical grounds of the perturbation theory successfully used for
description of the region of large  $Q^2$ (short distances
physics), contains the nonperturbative terms, determining the main
features of QCD in low  $Q^2$ region (long distances physics)
without the dramatic changes of these qualitatively different
regimes.

The analytic approach in QFT was formulated late in the 50s in
Refs.~\onlinecite{Red} and~\onlinecite{Bog} for QED and other
theories. For QCD  the analytic approach was applied
in~\cite{Shir,Doksh}. Indicate
Refs.~\onlinecite{SolShirTMF,Shir01,Magr,Pivovarov92,Stefanis,
Aguilar,Cvetic,Nest} in which the analytic approach and its
applications to QCD are considering.  The analyticity requirement,
which  follows from the general principles of QFT,  enables one to
resolve  the difficulties connected with nonphysical singularities
of the perturbation theory in the infrared region. In the
``analytically improved'' running coupling these singularities are
cancelled out by the nonperturbative contributions. In the
ultraviolet region the nonperturbative contributions  rapidly
decrease and the perturbative contributions are decisive.
Nonetheless, the behavior of the nonperturbative contributions in
the ultraviolet region originating from the procedure is of a
considerable interest. Their behavior appears to be important in
the construction of the synthetic running coupling.

Whereas for the one-loop case a separation  of the analytic
running coupling into the perturbative and the nonperturbative
components and behavior of the nonperturbative component are
obvious,  for the multi-loop analytic running coupling this is not
the case. For the two-loop case such separation in an explicit
form and study of the nonperturbative component of the analytic
running coupling was made in Ref.~\onlinecite{A}, for the
three-loop case it was made in~\cite{IHEP40,YadFiz}, and for the
four-loop case it was made in~\cite{JofP,FBS_03}. The
nonperturbative contributions in the analytic running coupling
were extracted explicitly and their expansion in powers of
$\Lambda^2/Q^2$ was obtained. The effective method of the precise
calculation of the analytic running coupling was developed on a
basis of this expansion.

It is known~\cite{Shir} that the analytic running coupling is
finite at zero. In Refs.~\onlinecite{JofP} and \onlinecite{FBS_03}
it was shown that the finiteness of the analytic  coupling at zero
was a consequence of the asymptotic freedom property of the
initial perturbation theory,
$\alpha_{\mathrm{an}}^\mathrm{n-loop}(0)=4\pi/b_0\simeq$ 1.396.
The running coupling is finite at zero also in the case of
``freezing'' of the interaction~\cite{Simonov}. The loss of
$Q^2$-dependence at low $Q^2$ takes place for a number of
different definitions of the running coupling~\cite{Deur}.
However, with such infrared behavior of the running coupling the
description  of the confinement and the dynamical breaking of the
chiral symmetry is not immediate. The behavior of the running
coupling corresponding to the infrared enhancement of the
interaction seems to be appropriate for description of these  most
important properties of QCD. In general the running coupling
constant is ambiguously determined. In the perturbation theory it
depends on the renormalization scheme choice. Besides it can
depend on the nonperturbative contributions in an observable used
for definition of the running coupling. In the so-called
nonperturbative V-scheme for the running coupling the singular
behavior $\alpha_V\sim 1/Q^2$ at $Q^2\rightarrow 0$  corresponds
to the linear confining quark-antiquark static potential at large
distances  with the universal string tension parameter  $\sigma$.
Point at Ref.~\onlinecite{tHooft} where the behavior of the
running coupling of the form $\sim 1/Q^2$ for all $Q^2$ takes
place for a perturbative treatment of theories with permanent
confinement. In QCD the static potential is defined in an explicit
gauge invariant form through the vacuum expectation value of the
Wilson loop. Its calculation on the lattice in quenched
approximation at large distances $r$ completely corresponds to the
string picture of heavy quarks
interaction~\cite{Greensite,Luscher} with confining term $\sigma
r$ and  a $\gamma/r$ correction  with a coefficient $\gamma$ as
predicted by the bosonic string theory.
 The synthetic running coupling under
consideration also belongs to the singular type couplings,
$\alpha_\mathrm{syn}\sim 1/Q^2$ at $Q^2\rightarrow 0$, and yet it
has its own  motivation related to the study of the
nonperturbative contributions  at $Q^2\rightarrow \infty$. The
synthetic running coupling appears practically to
coincide~\cite{AlekseevTMF05} with the initial perturbation theory
running coupling (we use $\alpha_s$ in the $\overline{MS}$-scheme)
in the region of its applicability but is well defined for all
$Q^2>0$.

The main methods of nonperturbative study of the Green's functions
in QCD and the running coupling which can be built out of these
functions  are solving the Dyson--Schwinger (DS) equations and
lattice calculations of the functional integrals. In
Ref.~\onlinecite{ShirIR} a summary of the results of such recent
studies is given, which can be supplemented with
papers~\cite{Nesterenko,Gogohia} where analytic methods were used
and~\cite{Burgio_et_al} with the lattice stimulation results. The
variety of the results for the behavior of $\alpha_s(Q^2)$ in the
infrared region is connected in particular with  different
truncation methods  applied  to close the DS equations. Besides,
solving  the closed integral equations (or systems  of equations)
requires, as a rule, the simplifying assumptions quite often
breaking the gauge symmetry and the pure technical approximations.
It is not surprising, than, that the results of the infrared
behavior study of $\alpha_s$ differ greatly from one another and
should not be compared literally. Let us note the review
papers~\cite{AnnPhys,ProgPart,PhysRept} of the investigations on
the IR behavior of the Green's functions, the running coupling in
QCD, and their applications in the hadron physics.

The possibility of the singular behavior $\alpha_s\sim 1/Q^2$ at
$Q^2\rightarrow 0$ which we consider has been  studied in a number
of papers. In particular, in Ref.~\onlinecite{Alek1} the infrared
behavior of the gluon Green's functions was studied by the
analytical calculations of the corresponding Feynman integrals of
the DS equation for the gluon propagator in the ghost-free axial
gauge where the running coupling was defined by the full gluon
propagator.  It was shown that the singular behavior of the gluon
propagator of the form $D(Q)\sim 1/(Q^2)^2$ at $Q^2\rightarrow 0$
was possible but it is essential to give up the commonly  used
approximation of the three-gluon vertex function by its
longitudinal part and  to take into account the transverse part of
the three-gluon vertex function of a definite form.

This paper is organized as follows. In Section~\ref{sec2} the
one-loop synthetic running coupling model of QCD for all $Q^2$ is
considered. In Section~\ref{sec3} we study  the analytic running
coupling for the standard  perturbation theory approximations up
to four-loop order and its separation into the perturbative and
the nonperturbative components. In Section~\ref{sec4} the
synthetic running coupling with the nonperturbative contributions
suppressed at large $Q^2$ is build on the basis of the analytic
running coupling. Setting the  parameters of the synthetic running
coupling is made. In the concluding section  the main results are
summarized and some remarks  made.

\section{\label{sec2}One-loop synthetic running coupling of QCD }

Let us consider the following additive modification of the
one-loop running coupling of QCD by means of the nonperturbative
pole type terms\footnote{This one-loop model was considered in
Refs.~\onlinecite{Samara97,YaF1998,ModPL1998} and finally
formulated with determination of the parameters from the energy
considerations in Ref.~\onlinecite{OneLoopModel1998}.}
\begin{eqnarray}
\alpha_{\mathrm{syn}}^{(1)}(Q^2)&=&\frac{4\pi}{b_0}\left[\frac{1}{
\ln(Q^2/\Lambda^2)}+\frac{\Lambda^2}{\Lambda^2-Q^2}
\right.\nonumber\\
&+&\left.\frac{c\Lambda^2}{Q^2}+\frac{(1-c)\Lambda^2}{Q^2+m^2_g}
\right],
\label{1}
\end{eqnarray}
where the gluon mass parameter
\begin{equation}
m_g=\frac{\Lambda}{\sqrt{c-1}}. \label{2}
\end{equation}
Here $Q^2$ is the Euclidean momentum squared, constant
$b_0=11-2n_f/3$ ($n_f$ is the number of active quark flavors),
$\Lambda$ is the dimensional parameter of the one-loop
model~(\ref{1}), $c$ is the dimensionless parameter of this model
(it is convenient to introduce the dimensional parameter
$\Lambda_1=\sqrt{c}\Lambda)$, $c\in(1,+\infty)$. The parameter
$\Lambda$ can be fixed, for example, by the normalization
condition at large  $Q^2$, whereas the parameter $c$ of the model,
as can be seen further, describes  the relation between the
parameter $\Lambda_{\mathrm{QCD}}$ and the string tension
parameter $\sigma$ of the string models\footnote{The
approximations of  QCD which take no account of the effects
connected with  masses of the heavy quarks contain only one
dimensional parameter. For example, it is $\Lambda_\mathrm{QCD}$
for large $Q^2$ or the string tension parameter $\sigma$ which is
adequate for low  $Q^2$. In the potential models the connection of
this parameters is studied in description of the bound states of
the heavy quarks~\cite{Buchmuller,Kiselev,Likhoded}.}. It stands
to reason that for the realistic definition of the parameters
$\Lambda_{\mathrm{QCD}}$, $\sigma$ and their connection it is
necessary to go out of the one-loop approximation.

The first term of Eq.~(\ref{1})  is the solution of the
renormalization group  equation  for the QCD running coupling
$\alpha_s{(Q^2)}$
\begin{equation}
Q^2\frac{\partial\alpha_s{(Q^2)}}{\partial Q^2}=\beta(\alpha_s)
\label{3}
\end{equation}
in the one-loop approximation,
$\beta(\alpha_s)\simeq-\beta_0\alpha_s^2$, $\beta_0=b_0/4\pi$.
Introducing the renormalization invariant parameter $\Lambda$ (the
integration constant of the differential equation) we obtain
\begin{equation}
\alpha^{(1)}_s(Q^2)=\frac{4\pi}{b_0}\frac{1}{\ln(Q^2/ \Lambda^2)},
\label{4}
\end{equation}
where a nonphysical singularity (the Landau pole) at
$Q^2=\Lambda^2$ is present.  The vanishing of expression~(\ref{4})
for $Q^2\rightarrow\infty$ corresponds to the remarkable property
of asymptotic freedom~\cite{Gross} of non-Abelian gauge theories,
while the growth of $\alpha_s$ (to some critical value or to
infinity) with decreasing $Q^2$  can be connected with the
confinement problem.

The pole terms in Eq.~(\ref{1}) are  nonperturbative,
$\Lambda^2\simeq\mu^2 \exp{\{-4\pi /b_0\alpha_s(\mu^2)\}}$. The
sum of the first two terms in Eq.~(\ref{1}) is an analytic
function in the complex $Q^2$-plane  with a cut from 0 to
$-\infty$,
\begin{equation}
\alpha^{(1)}_{\mathrm{an}}(Q^2)=\frac{4\pi}{b_0}\left[\frac{1}{\ln(Q^
2/ \Lambda^2)}+\frac{\Lambda^2}{\Lambda^2-Q^2}\right]. \label{5}
\end{equation}
This function can be presented by the dispersion relation
\begin{equation}
\alpha^{(1)}_\mathrm{an}(Q^2)=\int\limits_0\limits^{\infty}
\frac{d\sigma\,\tilde\rho^{(1)}(\sigma)}{\sigma+Q^2}, \label{6}
\end{equation}
where the function $\tilde\rho^{(1)}(\sigma)$ is called the
one-loop spectral density
\begin{equation}
\tilde\rho^{(1)}(\sigma)=\frac{4\pi}{b_0}\frac{1}{\ln^2(\sigma/
\Lambda^2)+\pi^2}. \label{7}
\end{equation}
The last equation can be obtained by the analytic continuation of
$\alpha_s$ into the Minkowski space $Q^2\rightarrow-\sigma-i0$ and
calculation of the imaginary part
$\tilde\rho^{(1)}(\sigma)=\frac{1}{\pi}\Im\alpha^{(1)}_
s(-\sigma-i0)$. For real $Q^2>0$ function~(\ref{5}) is positive
monotone decreasing function with maximum at zero
$\alpha_\mathrm{an}^{(1)}(0)=4\pi/b_0$. The second nonperturbative
term in Eq.~(\ref{5}) does not contribute to the imaginary part of
$\alpha_\mathrm{an}^{(1)}(Q^2)$ in going to the Minkowski space,
so that  $\tilde\rho^{(1)}(\sigma)= \frac{1}{\pi}\Im\alpha^
{(1)}_{\mathrm{an}}(-\sigma-i0)$. The synthetic running
coupling~(\ref{1}) can also be  presented in the form of the
dispersion relation
\begin{equation}
\alpha^{(1)}_\mathrm{syn}(Q^2)=\int\limits_{-0}\limits^{\infty}
\frac{d\sigma\,\tilde\rho_\mathrm{syn}^{(1)}(\sigma)}{\sigma+Q^2}.
\label{8}
\end{equation}
The function $\tilde\rho_\mathrm{syn}^{(1)}(\sigma)$ will be
called the one-loop spectral density for the synthetic running
coupling. It contains  additional terms in the form of the delta
functions,
\begin{equation}
\tilde\rho_\mathrm{syn}^{(1)}(\sigma)=\tilde\rho^{(1)}(\sigma)+
\frac{4\pi}{b_0}\left[c
\Lambda^2\delta(\sigma)+(1-c)\Lambda^2\delta(\sigma-m^2_g)\right].
\label{9}
\end{equation}
Introducing  two pole terms at $Q^2=0$ and $Q^2=-m_g^2<0$ does not
change the analyticity domain of  the analytic running
coupling~(\ref{5}). In Eq.~(\ref{9}) for the spectral density  the
additional  terms in the form of two $\delta$-functions localized
at $\sigma=0$ and $\sigma=m_g^2>0$ emerged (in comparison with
expression~(\ref{7}) of the perturbative origin).

Let us bring equation~(\ref{1}) for
$\alpha^{(1)}_\mathrm{syn}(Q^2)$ to the explicit renormalization
invariant form. It can be done without solving the differential
renormalization group equations. Writing the normalization
condition for $\alpha^{(1)}_\mathrm{syn}(Q^2)$ we obtain an
equation for the required dependence of the parameter
 $\Lambda^2$ on the values
$\alpha^{(1)}_\mathrm{syn}(\mu^2)$ and $\mu^2$ of the form
$$
\alpha^{(1)}_\mathrm{syn}(\mu^2) =\frac{4\pi}{b_0} \left [
\frac{1}{\ln (\mu^2/\Lambda^2)}+\frac{\Lambda^2}{\Lambda^2-\mu^2}+
\frac{c\Lambda^2}{\mu^2}\right.
$$
\begin{equation}
\left.+\frac{(1-c)\Lambda^2}{\mu^2+m^2_g}
\right ]. \label{10}
\end{equation}
From dimensional considerations
\begin{equation}
\Lambda^2=\mu^2 \exp\{-\varphi \left(a(\mu^2)\right)\}, \label{11}
\end{equation}
where $a(\mu^2)=(b_0/4\pi)\alpha^{(1)}_\mathrm{syn}(\mu^2)$. Then
for  $\varphi (a)$ we have a transcendental equation
\begin{equation}
a=\frac{1}{\varphi (a)}+\frac{1}{1-e^{\varphi (a)}}+ce^{-\varphi
(a)} - \frac{(c-1)^2}{1+(c-1)e^{\varphi (a)}}. \label{12}
\end{equation}
The function $\varphi (a)$   has the behavior $\varphi (a)\simeq
1/a\rightarrow+ \infty$ as $a\rightarrow +0$ for all values of
$c$. This behavior corresponds to the perturbative region. The
behavior of this solution at  $a\to +\infty$ is $\varphi (a)\simeq
-\ln(a/c)\rightarrow-\infty$. The beta function
$\beta_\mathrm{syn}(\alpha_\mathrm{syn})$ for the synthetic
running coupling can be found by  the equation which is analogous
to the equation~(\ref{3}),
\begin{equation}
Q^2\frac{\partial\alpha_\mathrm{syn}{(Q^2)}}{\partial
Q^2}=\beta_\mathrm{syn}(\alpha_\mathrm{syn}). \label{13}
\end{equation}
Differentiating the running coupling~(\ref{1}) with the use of
equations~(\ref{11}) and (\ref{12}), we obtain
\begin{eqnarray}
\beta_\mathrm{syn}(\alpha_\mathrm{syn})&=& \frac{4\pi}{b_0} \left
\{-a+\frac{1}{\varphi(a)}- \frac{1}{\varphi^2(a)}+
\frac{1}{(1-e^{\varphi(a)})^2}\right.
\nonumber\\
&-&\left.\frac{(c-1)^2}{(1+(c-1)e^{\varphi(a)})^2}\right \}
\Bigg\vert _{a=b_0\alpha_\mathrm{syn}/4\pi}. \label{14}
\end{eqnarray}
Therefore, using the behavior of  function $\varphi(a)$ as $a
 \rightarrow 0 $ and $ \infty$,
we find the asymptotic behavior
\begin{equation}
\beta_\mathrm{syn}(\alpha_\mathrm{syn})\simeq-\frac{b_0}{4\pi}\alpha_
\mathrm{syn}^2+ o(\alpha_\mathrm{syn}^2), \ \
\alpha_\mathrm{syn}\rightarrow0, \label{15}
\end{equation}
\begin{equation}
\beta_\mathrm{syn}(\alpha_\mathrm{syn})\simeq-\alpha_\mathrm{syn}-
\frac{4\pi}{b_0}c(c-2) +o(1), \ \ \alpha_\mathrm{syn}\rightarrow
\infty. \label{16}
\end{equation}
Doing the corresponding expansions we make sure that the
singularity of the $\beta$-function~(\ref{14})  at $\varphi
\rightarrow 0$ is seeming. We also make sure that for all
$\alpha_\mathrm{syn}>0$ the function
$\beta_\mathrm{syn}(\alpha_\mathrm{syn})$ is negatively defined.

Let us write the last three terms of the synthetic running
coupling~(\ref{1}), taking into account~(\ref{2}), in the form
$$
\alpha_{\mathrm{syn}}^{\mathrm{npt} \, (1)}(Q^2)=\frac{4\pi}{b_0}
\left[\frac{\Lambda^2}{\Lambda^2-Q^2}
+\frac{c\Lambda^2}{Q^2}+\frac{(1-c)\Lambda^2}{Q^2+m^2_g}\right]
$$
\begin{equation}
=\frac{4\pi}{b_0}\frac{c\Lambda^6}{Q^2(\Lambda^2-Q^2)(\Lambda^2+
(c-1)Q^2)}. \label{17}
\end{equation}
For $a\rightarrow +0$, according to equation~(\ref{12}) the
function $\phi(a) \simeq 1/a$, thus  for $\Lambda^2$ from
equation~(\ref{11}) we obtain  $\Lambda^2\simeq$
$\mu^2\exp{(-1/a)}$ at
$\alpha_{\mathrm{syn}}(\mu^2)=(4\pi/b_0)a\rightarrow +0$ so
expression~(\ref{17}) must be considered as a nonperturbative
component of the synthetic running coupling. The behavior of the
nonperturbative ``tail'' at large $Q^2$ is the following:
\begin{equation}
\alpha_{\mathrm{syn}}^{\mathrm{npt} \, (1)}(Q^2)=-\frac{4\pi}{b_0}
\left[\frac{c}{c-1}\frac{\Lambda^6}{Q^6}
\right]+O\left(\frac{\Lambda^8}{Q^8}\right). \label{19}
\end{equation}
As seen from Eq.~(\ref{19}), the nonperturbative contributions of
the synthetic running coupling  decreases at large $Q^2$
substantively faster than that of the analytic running
coupling~(\ref{5}).

Hence the one-loop synthetic running coupling of QCD has the
following interesting properties:
\begin{itemize}
\item[(i)] By the construction, as a function of  $Q^2$, it has an
analytic structure corresponding to the causality; that is, it is
a holomorphic  function in the complex $Q^2$-plane with a cut
along the negative real semiaxis. \item[(ii)] As a function of its
value $\alpha_{\mathrm{syn}}(\mu^2)$ at the normalization point
$\mu^2$, it has an essential singularity at the origin; the
asymptotic expansion of its nonperturbative part in
$\alpha_{\mathrm{syn}}(\mu^2)$ for $\alpha_{\mathrm{syn}}(\mu^2)
\rightarrow +0$ is equal to zero, which ensures  conformity  to
the initial perturbation theory. \item[(iii)] In the ultraviolet
region, it coincides with usual result of  the perturbation theory
(with renormalization invariance taken into account) apart from
fast decreasing power terms. The nonperturbative component behaves
as   $\sim 1/(Q^2)^ 3$ when $Q^2\rightarrow \infty$. \item[(iv)]
In the infrared region the synthetic running coupling does not
have nonphysical singularities of the perturbation theory. There
is the mass term and the singular at the origin  term which can be
responsible for the confinement of quarks.
\end{itemize}

As it will be evident from the subsequent considerations, all
these properties are valid for the two-, three- and four-loop
synthetic running coupling. It is significant that for the
one-loop case we have not only  representation~(\ref{8}), but for
one thing, the nonperturbative contributions are extracted from
the synthetic running coupling in the explicit form
\begin{equation}
\alpha_{\mathrm{syn}}^{(1)}(Q^2)=\alpha^{\mathrm{pt} \, (1)}(Q^2)+
\alpha_{\mathrm{syn}}^{\mathrm{npt}\,(1)}(Q^2) \label{20}
\end{equation}
and for another, for the nonperturbative contributions, a simple
formula  is on hand.  For large $Q^2$ ($Q^2>\Lambda^2$) this
contributions can be represented as a series
\begin{equation}
\alpha_{\mathrm{syn}}^{\mathrm{npt} \, (1)}(Q^2)=-\frac{4\pi}{b_0}
\left(\frac{\Lambda^2}{Q^2}\right)^3\sum\limits^ {\infty}\limits_
{n=0}\left(1-(1-c)^{-n-1}\right)\left(\frac{\Lambda^2}{Q^2}\right)^n.
\label{21}
\end{equation}
For small $Q^2$ ($Q^2<\Lambda^2$) we have the expansion
\begin{equation} \alpha_{\mathrm{syn}}^{\mathrm{npt} \,
(1)}(Q^2)=\frac{4\pi}{b_0} \frac{\Lambda^2}{Q^2}\sum\limits^
{\infty}\limits_
{n=0}\left(1-(1-c)^{n+1}\right)\left(\frac{Q^2}{\Lambda^2}\right)^n.
\label{22}
\end{equation}

The case $c=2$ is the particular one from the symmetry
considerations,
$$
\alpha_{\mathrm{syn}}^{\mathrm{npt} \, (1)}(Q^2)=\frac{4\pi}{b_0}
\left[ \frac{\Lambda^2}{\Lambda^2-Q^2}
+\frac{2\Lambda^2}{Q^2}-\frac{\Lambda^2}{\Lambda^2+Q^2}\right]
$$
\begin{equation}
=\frac{4\pi}{b_0}\frac{2\Lambda^2}{Q^2}\frac{\Lambda^4}{\Lambda^4-
Q^4}, \label{23}
\end{equation}
for which the nonperturbative component is the odd function of
$Q^2$,
\begin{equation}
\alpha_{\mathrm{syn}}^{\mathrm{npt} \, (1)}(-Q^2)=
-\alpha_{\mathrm{syn}}^{\mathrm{npt} \, (1)}(Q^2). \label{24}
\end{equation}
Here, in the ultraviolet expansion as well as in the infrared
expansion there are no terms of the even powers of $Q^2$. In
particular, the infrared expansion does not have the Coulomb's
mode.

Indicate for completeness  two boundary cases for the values
$c\in(1,+\infty)$ considered. The first case is $c=1$, for which
$$
\alpha_{\mathrm{syn}}^{\mathrm{npt} \, (1)}(Q^2)=
\frac{4\pi}{b_0}\frac{\Lambda^4}{Q^2(\Lambda^2- Q^2)},
$$
and the nonperturbative component decreases at infinity not so
fast as in expression~(\ref{19}). The second one is  $c=\infty$
for which
$$
\alpha_{\mathrm{syn}}^{\mathrm{npt} \, (1)}(Q^2)=
\frac{4\pi}{b_0}\frac{\Lambda^6}{Q^4(\Lambda^2- Q^2)},
$$
and the singularity in the infrared region is stronger, $
\alpha_{\mathrm{syn}}^{(1)}(Q^2)\sim 1/(Q^2)^2$, $Q^2\rightarrow
0$.

\section{\label{sec3}Multi-loop analytic running coupling of QCD}

For the multi-loop case the renormalization group
equation~(\ref{3}) for the QCD running coupling $\alpha_s(Q^2)$ is
of the form
$$
Q^2\frac{\partial\alpha_s(Q^2)}{\partial Q^2}=
\beta(\alpha_s)
$$
\begin{equation}
=-\beta_0\alpha_s^2-\beta_1\alpha_s^3-
\beta_2\alpha_s^4 -\beta_3\alpha_s^5+O(\alpha_s^6). \label{25}
\end{equation}
The coefficients  $\beta_0$, $\beta_1$ do not depend on the
renormalization scheme choice, whereas the next coefficients do
depend on this choice. For the numerical calculations we use its
values within the   $\overline{MS}$-scheme.

Let us write down the solution of equation~(\ref{25}) for
$\alpha_s(Q^2)$ at $L=\ln(Q^2/\Lambda^2)\rightarrow\infty$ in the
form of the standard expansion in inverse powers of the logarithms
$$
\alpha_s(Q^2)=\frac{1}{\beta_0 L}\left\{1-\frac{\beta_1}{\beta_0
^2L}\ln L\right.
$$
$$
+\frac{\beta_1^2}{\beta_0^4L^2} \left[\ln^2 L-\ln
L-1+\frac{\beta_0 \beta_2}{\beta_1^2}\right]
$$
$$
-\frac{\beta_1^3}{\beta_0^6L^3}\left[ \ln^3 L -\frac{5}{2}\ln^2
L-\left(2-\frac{3\beta_0\beta_2} {\beta_1^2}\right)\ln L\right.
$$
\begin{equation}
\left.\left.+\frac{1}{2}-\frac{\beta_0^2
\beta_3}{2\beta_1^3}\right]+O\left(\frac{1}{L^4}\right)\right\}.
\label{27}
\end{equation}
The sum of the terms of Eq.~(\ref{27}) up to   $1/L^n$ order
(n=1,2,3,4) will be referred to further on by  the $n$-loop
perturbative component of the running coupling  and denoted as
$\alpha^\mathrm{pt}(Q^2)$. It can be written in the form
\begin{equation}
\alpha^\mathrm{pt}(Q^2) =\frac{4\pi}{b_0}a(x), \label{28}
\end{equation}
$$
a(x)=\frac{1}{\ln x}- b\frac{\ln(\ln
x)}{\ln^2x}+b^2\left[\frac{\ln^2(\ln x)}{\ln^3x}- \frac{\ln(\ln
x)}{\ln^3x}+\frac{\kappa}{\ln^3x}\right]
$$
\begin{equation}
-b^3\left[\frac{\ln^3(\ln x)}{\ln^4x}-\frac{5}{2}\frac{\ln^2(\ln
x)}{\ln^4x}+(3\kappa+1) \frac{\ln(\ln
x)}{\ln^4x}+\frac{\bar\kappa}{\ln^4x}\right]. \label{29}
\end{equation}
Here $x=Q^2/\Lambda^2$ and the coefficients
\begin{equation}
b=\frac{\beta_1}{\beta^2_0}, \ \ \ \kappa=-1+\frac{\beta_0
\beta_2}{\beta_1^2}, \ \ \
\bar\kappa=\frac{1}{2}-\frac{\beta_0^2\beta_3}{2\beta_1^3}.
\label{30}
\end{equation}
The values of $b_0$, $b$, $\kappa$, $\bar\kappa$ depend on $n_f$.
Within the standard picture of matching the solutions at heavy
quark thresholds the parameter   $\Lambda$ becomes dependent on
$n_f$\footnote{In perturbation theory the parameter  $\Lambda$
depends also on the renormalization scheme choice. To study the
deviation of the nonperturbative couplings and their consequences
from the corresponding perturbative quantities the choice of the
standard renormalization scheme $\overline{MS}$ is convenient.}.

The analytic running coupling $\alpha_\mathrm{an}(Q^2)
=(4\pi/b_0)a_\mathrm{an}(x)$ is defined through  the dispersion
relation
\begin{equation}
a_{\mathrm{an}}(x)=\frac{1}{\pi}\int\limits_0^\infty
\frac{d\sigma}{x+\sigma} \rho(\sigma), \label{31}
\end{equation}
where the spectral density $\rho(\sigma)=\Im
a_{\mathrm{an}}(-\sigma-i 0)$.  The analytic approach
suggests\footnote{This is precisely the variant of the ``analytic
improvement'' procedure    which we consider. The running coupling
obtained in this way we call the analytic running coupling.} that
 $\Im a_{\mathrm{an}}(-\sigma-i 0)=\Im
a(-\sigma-i 0)$. As a result from  function  $a(x)$ of the
form~(\ref{29}) with nonphysical singularities on the positive
real semiaxis of the complex plane $x=Q^2/\Lambda^2$ we come to
the function $a_\mathrm{an}(x)$ of the form~(\ref{31}), which is a
single-valued analytic function  in the complex plane  $x$ with a
cut from 0 to $-\infty$ (with a standard definition of the cuts of
the logarithmic function).  In Refs.~\onlinecite{JofP,FBS_03} up
to four-loop order the separation of the analytic running coupling
into the perturbative and  nonperturbative components  was
obtained,
\begin{equation}
\alpha_{\mathrm{an}}(Q^2)=\alpha^{\mathrm{pt}}(Q^2)+
\alpha^{\mathrm{npt}}_{\mathrm{an}}(Q^2). \label{32}
\end{equation}
In Eq.~(\ref{32}) we take  $\alpha^ {\mathrm{pt}}(Q^2)$ as the
initial (in our case the standard) solution of the renormalization
group equation up to four-loop order and
$\alpha^{\mathrm{npt}}_{\mathrm{an}}(Q^2)$ appears to arise
additionally as a result of the  procedure.  The following
expansion was obtained in the power series
\begin{equation}
\alpha^{\mathrm{npt}}_{\mathrm{an}}(Q^2)=\frac{4\pi}{b_0}
\sum\limits^{\infty }\limits_
{n=1}c_n\left(\frac{\Lambda^2}{Q^2}\right)^n, \label{33}
\end{equation}
where the coefficients $c_n$ were  defined by the beta function
coefficients. Note the important properties of
expansion~(\ref{33}) such as the higher loop stability of the
coefficient of the leading term and the slow increase of the
coefficients $c_n$ with number  $n$.

\section{\label{sec4}Multi-loop synthetic running coupling of QCD}
The one-loop synthetic running coupling can be naturally extended
to the multi-loop cases. Thus modify the analytic running coupling
introducing two additional nonperturbative terms, the singular at
zero term of the form $\sim 1/Q^2$ and the mass term of the form
$\sim 1/(Q^2+m_g^2)$. As a result,  we come to the expression
\begin{equation}
\alpha_{\mathrm{syn}}(Q^2)= \alpha_{\mathrm{an}}(Q^2)
+\frac{4\pi}{b_0}\left[ \frac{c\Lambda^2}{Q^2}-\frac{d\Lambda^2}
{Q^2+m_g^2}\right], \label{35}
\end{equation}
containing, besides   $\Lambda$, three  subsidiary parameters:
$c$, $d$ and $m_g$ ($m_g\equiv m_\Lambda \Lambda$), which are
assumed to be nonzero.  We will define this parameters in the
following way. At large  $Q^2$ $(Q^2>\Lambda^2)$, using
expansion~(\ref{33}), we can find
$$
\alpha_{\mathrm{syn}}(Q^2)=\alpha^{\mathrm{pt}}(Q^2)+
\frac{4\pi}{b_0}\left[(c_1+c-d)
\frac{\Lambda^2}{Q^2}\right.
$$
$$
+(c_2+dm_\Lambda^2)\left(\frac{\Lambda^2}
{Q^2}\right)^2\nonumber\\
+\left.(c_3-dm_\Lambda^4)\left(\frac{\Lambda^2}{Q^2}\right)^3
\right]
$$
\begin{equation}
+O\left(\left(\frac{\Lambda^2}{Q^2}\right)^4\right).
\label{36}
\end{equation}
Demand the nonperturbative contributions to be minimal at large
$Q^2$, i.e.  the terms of the form $\sim 1/Q^2$, $\sim 1/(Q^2)^2$
to be absent in Eq.~(\ref{36})\footnote{This condition corresponds
to the principle of minimality of the nonperturbative
contributions in the perturbative ultraviolet region~\cite{
YaF1998,ModPL1998}. Other reasoning is to be used to define the
parameters in  Eq.~(\ref{35}) if one builds the effective charge
for observable with power terms at large $Q^2$ (see e.g.,
Ref.~\onlinecite{Chernodub} for $1/Q^2$ power corrections in
$\alpha_{\mathrm{V}}(Q^2)$).}. Then two of three parameters are
fixed by the following equations
\begin{equation}
d=c+c_1, \,\,\, m_\Lambda^2=-c_2/(c+c_1). \label{37}
\end{equation}
The parameter $\Lambda_1=\sqrt{c}\Lambda$  will be considered as
fixed. The coefficients   $c_n<0$, therefore the absence of
tachion condition is $\Lambda<\Lambda_1/\sqrt{-c_1}$. With a given
number of loops the free parameter of the synthetic running
coupling is only one parameter $\Lambda$. Then taking into account
Eq.~(\ref{37}) we have
$$
\alpha_{\mathrm{syn}}(Q^2)=\alpha^{\mathrm{pt}}(Q^2)+
\frac{4\pi}{b_0}\left[c_3-
\frac{c_2^2\Lambda^2}{\Lambda_1^2+c_1\Lambda^2} \right]
\left(\frac{\Lambda^2}{Q^2}\right)^3
$$
\begin{equation}
+O\left(\left(Q^2\right)^{-4}\right). \label{38}
\end{equation}
As seen from Eq.~(\ref{38}), the leading power nonperturbative
term decreases rapidly at  $Q^2\rightarrow\infty$ and in the
absence of the tachion is negative\footnote{The nonperturbative
``tail''  as a whole of the running coupling
$\alpha_{\mathrm{syn}}(Q^2)$ turns out to be negative at large
$Q^2$.}. For the gluon mass parameter $m_g$ which we call the
dynamical gluon mass from Eqs.~(\ref{37}) it follows that
\begin{eqnarray}
m_g=\Lambda\sqrt{\frac{-c_2\Lambda^2}{\Lambda_1^2+c_1\Lambda^2}}.
\label{39}
\end{eqnarray}
%

\begin{figure*}[tbh]
\includegraphics[width=10cm]{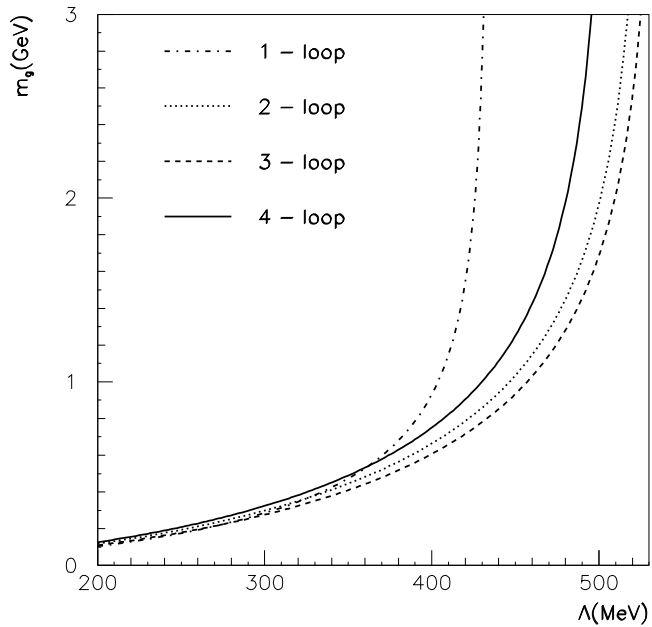}
\caption{The gluon mass parameter $m_g$ as a function of $\Lambda$
for different number of loops of the initial perturbation theory
approximation.\label{fig1}}
\end{figure*}
Let us turn to the interpretation of the parameter $\Lambda_1$
which describes the value of the singular term. As it has already
been noted,  the behavior of the running coupling  $\alpha_V \sim
1/Q^2$ at $Q^2\rightarrow 0$  corresponds to the linear confining
quark-antiquark static potential in quenched QCD.
We set up a correspondence of
the potential and the running coupling using the equation (to
compare with Refs.~\onlinecite{BogShir} and
~\onlinecite{Buchmuller})
\begin{equation}
V(r)=-\frac{4}{3}\int
\frac{d^nq}{(2\pi)^n}\exp{(i\mathbf{q}\mathbf{r})}\frac{4\pi
\alpha_\mathrm{V}(q^2)}{q^2}\Bigg\vert_{n=3}, \label{40}
\end{equation}
where $\alpha_\mathrm{V}(q^2)$ is defined as an effective
charge~\cite{Grunberg}, which is the renormalization scheme
independent and gauge invariant quantity. The color factor
corresponds to  the $SU(N_c)$ group, $N_c=3$. Let us assume that
in the infrared region
\begin{equation}
\alpha_\mathrm{V}(q^2)\simeq\frac{3}{2}\frac{\sigma}{q^2}, \ \ \
q^2\rightarrow 0. \label{41}
\end{equation}
Then the integral over three dimensional momentum space  in
Eq.~(\ref{40}) formally diverges at the origin. We define this
integral introducing the dimensional regularization. After
integration over the $n$-dimensional Euclidean momentum space we
put $n=3$.  Then  as the divergences do not occur, the transition
to the $n$-dimensional integration provides not only the
regularization but the definition of the divergent integral for
the case  $n=3$. As a result, for the infrared behavior of the
effective charge~(\ref{41}) the behavior of the potential at large
distances is as follows:
\begin{equation}
V(r)\simeq \sigma r,  \ \ \  r\rightarrow\infty, \label{42}
\end{equation}
where $\sigma\equiv a^2$ is the string tension parameter. Let us
define the parameter $\Lambda_1$ of the synthetic running coupling
$\alpha_\mathrm{syn}$ from the correspondence of the singular at
zero term in Eq.~(\ref{35}) to the infrared behavior~(\ref{41}) of
the running coupling $\alpha_\mathrm{V}$. Then
\begin{equation}
\frac{3}{2}\sigma=\frac{4\pi}{b_0}\Lambda_1^2,  \ \ \
\Lambda_1^2=c\Lambda^2. \label{43}
\end{equation}
Therefore, if the string tension parameter is given, the parameter
$\Lambda_1$ can be specified by Eq.~(\ref{43}). Then with $a\simeq
0.42$ GeV, $b_0=9$ we obtain\footnote{The string tension parameter
$\sigma=$ (0.42 GeV)$^2$  in particular, in the relativistic
string model with massive quarks at the ends of the
string~\cite{Soloviev98&00}. Then the slope of Regge trajectories
$\alpha'=1/(2\pi\sigma)\simeq 0.90$ GeV$^{-2}$.} $\Lambda_1\simeq$
435 MeV.  The parameter $\Lambda$ (as well as the parameter $c$)
can be fixed by  the normalization condition, and the synthetic
running coupling will be fixed completely.

Consider the dependence of the dynamical gluon mass on $\Lambda$
for different number of loops of the initial perturbation theory
approximation. In Fig.~\ref{fig1} the dynamical gluon mass
$m_g(\Lambda)$ is shown for  one --- four-loop cases. Up to 400
MeV the curves do not diverge too much and at $\Lambda=$ 375 MeV
$m_g\simeq$ 0.6 GeV.
\begin{table}[htb]
\caption{ The parameters  $\Lambda_\mathrm{pt}$ (MeV),
$\Lambda_\mathrm{an}$ (MeV), $\Lambda_\mathrm{syn}$ (MeV), the
dynamical gluon mass $m_g$ (MeV) and the parameters $c$, $d$ on
the number of loops. The number of active quark flavors $n_f=$ 3,
$\Lambda_1=$ 435 MeV. The normalization condition is
$\alpha(M_{\tau}^2)=$ 0.33, $M_{\tau}=$ 1.777 GeV.} \label{tab1}
\begin{ruledtabular}
\begin{tabular}{lrrrr}                 &1-loop
&2-loop&3-loop&4-loop
     \\\hline
$\Lambda_\mathrm{pt}$
 &       214.25 &       395.10 &       364.19 &       357.32
\\
$\Lambda_\mathrm{an}$
 &       254.51 &       636.02 &       523.86 &       535.54
\\
$\Lambda_\mathrm{syn}$
 &       214.26 &       397.10 &       365.26 &       358.70
\\
$m_g$
 &       121.14 &       648.09 &       461.66 &       526.57
\\
$c $
 &       4.1282 &       1.2018 &       1.4204 &       1.4728
\\
$d $
 &      3.1282 &      0.5358 &      0.7706 &      0.7447
\\
\end{tabular}
\end{ruledtabular}
\end{table}
We normalize the running couplings  $\alpha_{\mathrm{syn}}(Q^2)$,
$\alpha_{\mathrm{an}}(Q^2)$ and $\alpha^{\mathrm{pt}}(Q^2)$ at the
$\tau$-lepton mass by~\cite{Data,Piv} $ \alpha(M_\tau^2)=0.33,
\,\,\, M_\tau=1.777 $ GeV. For this normalization condition the
values of the parameters $\Lambda$, the dynamical gluon mass $m_g$
and the parameters $c$, $d$ are given in Table~\ref{tab1}. Point
to two things. The parameters  $\Lambda_\mathrm{syn}$ and
$\Lambda_\mathrm{pt}$ are close in value  whereas the parameters
$\Lambda_\mathrm{an}$ are considerably  larger. This is a
consequence of  conditions~(\ref{37}) which give the fast decrease
of the nonperturbative terms of $\alpha_\mathrm{syn}(Q^2)$ at
large $Q^2$. For all quantities considered the one-loop case turns
out to be exceptional, and then the stabilization is observed with
the number of loops of the initial perturbation theory
approximation.

\section{\label{sec5}Conclusions}
In the construction of the QCD analytic running
coupling~(\ref{31}) the nonphysical singularities of the
perturbation theory in the infrared region disappear and in the
ultraviolet region the nonperturbative power corrections arise
decreasing rapidly at large $Q^2$, in comparison with the main
perturbative component. However, when considering the
nonperturbative quantities it may happens  that  the decrease of
the nonperturbative contributions is not fast enough for the
consistent definition of these  quantities. In the synthetic
running coupling it is proposed to provide the highest possible
suppression of the nonperturbative contributions at large $Q^2$ by
means of a minimal number of the pole type terms. The parameters
characterizing the additional nonperturbative terms have a clear
physical meaning and take the reasonable values. The running
coupling~(\ref{1}) built on the basis of the analytic running
coupling~(\ref{5}) is called the one-loop synthetic running
coupling  because it contain the parameters related to the
ultraviolet region as well as to the infrared region. We introduce
the singular at zero term of the form $\sim 1/Q^2$ which can
correspond to the linear quark confinement and the mass term of
the form $\sim 1/(Q^2+m_g^2)$ with the parameter $m_g$
corresponding to  the non-vanishing dynamical gluon mass. We
impose the condition of the fastest decrease of the
nonperturbative component at large $Q^2$ and receive the running
coupling model of  form~(\ref{1}). The model
$\alpha_\mathrm{syn}^{(1)}(Q^2)$ has two independent parameters,
the dimensional parameter $\Lambda$ and the dimensionless
parameter $c$  which defines the value of the singular term. In
Section 2 the one-loop model of the synthetic running coupling and
its nonperturbative component properties for $c\in[1,+\infty)$ are
considered. A study of the multi-loop analytic running coupling
and its nonperturbative component provide a possibility of natural
generalization of the synthetic running coupling model  to the
multi-loop cases. The multi-loop synthetic running
coupling~(\ref{35}) as the one-loop model is constructed by
introducing two additional nonperturbative terms of the form $\sim
1/Q^2$ and $\sim 1/(Q^2+m_g^2)$. The minimality principle of the
nonperturbative contributions in the perturbative region leads to
two equations~(\ref{37}) for the introduced nonperturbative
parameters. As a result, the synthetic running coupling  has two
independent parameters. First, the parameter $\Lambda$ which owing
to highly fast decrease of the nonperturbative contributions at
large  $Q^2$ practically coincides with the parameter
$\Lambda_\mathrm{QCD}$ in the region of application  of the
perturbative solutions. Second,  the dimensionless parameter $c$
(or the dimensional parameter $\Lambda_1=\sqrt{c}\Lambda$)
determining the value of the singular term. Correlating this
parameter responsible for the infrared enhancement by
Eqs.~(\ref{43}) with the string tension parameter $\sigma$ of the
string models, we arrive at the dynamical gluon mass as a function
of the parameter  $\Lambda$  defined by Eq.~(\ref{39}). The
corresponding dependencies for 1--4-loop cases are given in
Fig.~\ref{fig1} with  $\Lambda_1=435$ MeV (that is
$\sigma^{1/2}=0.42$ GeV). The normalization completely defines the
synthetic running coupling and for $\alpha(M_\tau^2)=$ 0.33 the
values of the parameters are shown in Table~\ref{tab1}. The
parameter $m_g$ in Eq.~(\ref{39}) for
$\Lambda<\Lambda_1/\sqrt{-c_1}$ is real, and the synthetic running
coupling $\alpha_\mathrm{syn}(Q^2)$  in Eq.~(\ref{35}) is a
holomorphic function in the complex plane  $Q^2$ with a cut  along
the real negative semiaxis. Thus the synthetic running coupling of
QCD $\alpha_\mathrm{syn}(Q^2)$ has the properties outlined in
Section 2 for the one-loop synthetic running coupling model.

As  seen in Table~\ref{tab1} the parameters of the synthetic
running coupling fixed in such a way show the higher loop
stabilization.  In particular, the dynamical gluon mass $m_g$ can
be estimated as 400--600 MeV. For $\sigma=$ (0.42 GeV)$^2$ and
$\alpha_\mathrm{syn}(M^2_\tau)=$ 0.32, 0.33, 0.34  it is obtained
that $m_g=$ 453, 527, 613  MeV (for the 4-loop
case)~\footnote{Estimates of the value of the gluon mass can be
found in Ref.~\onlinecite{Field}. Our estimate $m_g=$ 530 $\pm 80$
MeV is in agreement for example with the phenomenological value of
the dynamical gluon mass $m_g\approx$ 400$^{+350}_{-100}$ MeV of
Ref.~\onlinecite{Luna}.}. Hence the string tension identification
of the parameter of the synthetic running coupling defining the
value of the singular term results in  the consistent values for
the other parameters considered.

According to Eq.~(\ref{38})  the nonperturbative contributions
decrease at large $Q^2$ as $\sim 1/(Q^2)^3$, which is sufficient
for the convergence of the gluon condensate in the ultraviolet
region~\cite{OneLoopModel1998}. Thereupon the generalization of
the gluon condensate studies~\cite{OneLoopModel1998,0411339} to
the multi-loop  synthetic running coupling  is of much interest.

\begin{acknowledgments}
I am indebted to  B.A.~Arbuzov, V.A.~Petrov, and V.E.~Rochev for
useful discussions. I would like to thank D.V.~Shirkov for helpful
advices and stimulating discussions. This work has been partly
supported by RFBR under Grant No.~05-01-00992.
\end{acknowledgments}

\end{document}